\documentclass[aps,prx,twocolumn,superscriptaddress,nofootinbib,longbibliography]{revtex4-1}

\usepackage{hyperref}
\usepackage{amsmath}
\usepackage{amssymb}
\usepackage{amsthm}
\usepackage{amsbsy}
\usepackage{mathrsfs}
\usepackage{graphicx}
\usepackage{enumerate}
\usepackage{array}
\usepackage{enumitem}
\usepackage{float}
\usepackage{comment}
\usepackage{soul}
\usepackage{epstopdf}

\theoremstyle{plain}
\newtheorem{thm}{Theorem}

\newtheorem{axiom}{Axiom}
\newtheorem{defin}{Definition}
\newtheorem{prin}{Principle}

\newcommand{\tsc}[1]{\textsc{#1}}

\usepackage{color}

\definecolor{red}{rgb}{1,0,0}
\definecolor{blue}{rgb}{0,0,0.8}
\definecolor{maroon}{rgb}{0.65,0,0}
\definecolor{ngreen}{rgb}{0.2,0.5,0.2}
\definecolor{pink}{rgb}{1.0,0.3,0.7}
\definecolor{golden}{rgb}{0.8,0.6,0.1}

\newcommand{\red}{\color{black}}
\newcommand{\blk}{\color{black}}

\newcommand{\RC}{\textsc{Relativistic Causality}}
\newcommand{\RCA}{\textsc{Relativistic Causal Arrow}}

\newcommand{\TCA}{\textsc{Temporal Causal Arrow}}

\newcommand{\ST}{\textsc{Space-Time}}
\newcommand{\DE}{\textsc{Decorrelating Explanation}}
\newcommand{\RPCC}{\textsc{Reichenbach's Principle of Common Cause}}
\newcommand{\PCC}{\textsc{Principle of Common Cause}}
\newcommand{\AOE}{\textsc{Absoluteness of Observed Events}}
\newcommand{\LOC}{\textsc{Locality}}
\newcommand{\NSD}{\textsc{No-Superdeterminism}}
\newcommand{\PRED}{\textsc{Predetermination}}
\newcommand{\LC}{\textsc{Local Causality}}
\newcommand{\II}{\textsc{Independent Interventions}}
\newcommand{\IC}{\textsc{Interventionist Causation}}
\newcommand{\LF}{\textsc{Local Friendliness}}
\newcommand{\LA}{\textsc{Local Agency}}
\newcommand{\NFT}{{No Fine-Tuning}}
\newcommand{\cause}{\textsc{cause}}
\newcommand{\causes}{\textsc{causes}}

\newcommand{\event}{\textsc{event}}
\newcommand{\events}{\textsc{events}}

\newcommand{\etal}{{\em et al.}}

\begin{document}

\title{Implications of Local Friendliness violation for quantum causality}

\author{Eric G. Cavalcanti}
\email{e.cavalcanti@griffith.edu.au}
\affiliation{Centre for Quantum Dynamics, Griffith University, Gold Coast, QLD 4222, Australia}

\author{Howard M. Wiseman}
\email{h.wiseman@griffith.edu.au}
\affiliation{Centre for Quantum Dynamics, Griffith University, Brisbane, QLD 4111, Australia}

\begin{abstract}
We provide a new formulation of the Local Friendliness no-go theorem of Bong \etal~[Nat. Phys. 16, 1199 (2020)] from fundamental causal principles, providing another perspective on how it puts strictly stronger bounds on quantum reality than Bell's theorem. In particular, quantum causal models have been proposed as a way to maintain a peaceful coexistence between quantum mechanics and relativistic causality, while respecting Leibniz's methodological principle.  This works for Bell's theorem but does not work for the Local Friendliness no-go theorem, which considers an extended Wigner's Friend scenario. More radical conceptual renewal is required; we suggest that cleaving to Leibniz's principle requires extending relativity to events themselves. 
\end{abstract}

\date{\today}

\maketitle


\section{Introduction}

\begin{quote}
    ``For me \dots\ this is the real problem with quantum theory: the apparently essential conflict between any sharp formulation [of quantum theory] and fundamental relativity \dots. It may be that a real synthesis of quantum and relativity theories requires not just technical developments but radical conceptual renewal''. (John S.~Bell, 1984 \cite{Bell1984}).
\end{quote}

In a recent work~\cite{CQD2020}, we (the present authors together with co-authors) proved what we called a ``strong'' no-go theorem on the Wigner's friend paradox---the ``Local Friendliness'' (LF) no-go theorem. In this paper, we discuss further in what sense it is stronger than Bell's theorem and give a broader picture of some of its implications. In particular, we discuss how it presents a challenge to a popular resolution of Bell's theorem that goes back to Shimony~\cite{Shimony1984}. 

Shimony proposed to separate Bell's notion {of} 
 \LC\ into two independent assumptions, which he called ``Parameter Independence'' (PI) and ``Outcome Independence'' (OI). His idea was that, while violation of PI was undoubtedly a case of action-at-a-distance---and was thus contrary to the letter of the theory of relativity---violation of OI was a far milder affliction---``passion at a distance''~\cite{Shimony1984}---that was contrary only to the spirit of the theory of relativity. This allowed, the argument goes, for a ``peaceful coexistence'' \cite{Shimony1984, Myrvold2002, Berkovitz2007} between quantum theory and relativity.  

Shimony's proposal has been criticized for missing the mark. For Bell, it left open the question of causal explanation of correlations \cite{Bell1990}. The program of quantum causal models~\cite{Leifer2013,Cavalcanti2014,Pienaar2015,Costa2016,Allen2017,Barrett2019} has provided a candidate answer to this challenge by generalising the classical framework of causal models \cite{Pearl2000} to accommodate quantum correlations. This program shares important similarities with Shimony's early proposal, and as we will show, is subject to the same challenge in light of the LF no-go theorem.

To better put these results and discussions in context, we present the Local Friendliness theorem within an updated version of the conceptual framework introduced by us \mbox{in \cite{Causarum},} where we reformulated the two theorems of John Bell (the 1964~\cite{Bell1964} and 1976~\cite{Bell1976}  theorems \red -- see Ref.~\cite{Wiseman2014} by one of us for a detailed discussion of the history and controversy surrounding the distinction.\blk) in terms of fundamental metaphysical principles concerning events, space-time, and causality. This work was partly motivated by disagreements between two broad interpretational camps we referred to (following Ref.~\cite{Wiseman2014}) as ``realists'' and ``operationalists''. \red (As an aside: some whose positions are not in what we call the ``realist'' camp would nevertheless claim to be realists in some sense. We will keep the scare quotes when referring to these two camps here to avoid debates about the meaning of those terms.) \blk  ~Many of these disagreements arose from (often implicit) assumptions about the \emph{{meaning}
} of terms like ``local'', and by using deeper principles, the foundations of these different perspectives could be made clearer. Moreover, we could present a ``conciliatory'' reformulation (Theorem 8 of \cite{Causarum}), in which the two camps could agree on the meaning of all the assumptions involved but only disagree about which assumptions were to be \emph{{rejected}} in light of \mbox{the theorem. }

As we will discuss in detail in this paper, the new challenge alluded to in the opening  paragraph here is a challenge for the operationalist camp. In our Bell-conciliation theorem~\cite{Causarum}, the ``realist'' camp would reject the principle {of} \RC, while the ``operationalist'' camp would reject  
that {of} \DE. This latter principle is the ``quantitative'' part of \RPCC~ (RPCC), which we split into two parts following the analysis proposed by EGC and Lal~\cite{Cavalcanti2014}. Rejection of this quantitative part of RPCC is also, implicitly or explicitly, the approach to resolve Bell's theorem taken within the frameworks of quantum causal \mbox{models~\cite{Pienaar2015,Costa2016,Allen2017,Barrett2019}.} However---and this is the new challenge---the assumption of \DE\ is \emph{not required} for the derivation of LF inequalities.

What are the implications of this? Firstly, there is of course the possibility that the LF inequalities derived in~\cite{CQD2020} fail to be violated with ``genuine'' observers (the ``friends'' of Wigner). We discuss this question at length in another {paper} 
\cite{LF_AI}. For simplicity, here, we just acknowledge this as an open question and consider the implications of taking seriously the possibility that the LF violations demonstrated for very simple ``observers'' in recent experiments~\cite{Proietti2019,CQD2020} can be maintained at the level of systems we may be strongly inclined to consider genuine observers---e.g., human-level artificial intelligences running in a very large quantum {computer}~\cite{LF_AI}.

Assuming that nature violates Local Friendliness, then, this implies that a popular class of resolutions of Bell's theorem---encompassing both ``passion at a distance'' and quantum causal models---is not available to resolve the LF theorem. Is a ``peaceful coexistence'' between quantum theory and relativity therefore impossible? 

A ``realist'' would say yes. This camp would see the LF no-go theorem as a strengthening of the ``nonlocal'' position (rejecting \RC), since it demonstrates more clearly than with Bell's theorem the depth of the ``essential conflict'' between quantum theory and relativity, making it sharper than ever the need for ``radical conceptual renewal'', in Bell's words.

On the other hand, an ``operationalist'' may also find hope in a different sort of radical conceptual renewal. We will argue that Leibniz's Principle of the Identity of Indiscernibles~\cite{Spekkens2019}---a methodological principle underlying Einstein's principles of relativity and of equivalence, as well as the program of quantum causal models---can only be maintained by rejecting \AOE. This last is one of the assumptions of the LF no-go theorem, as well as a (typically implicit) assumption in Bell's theorem. Maintaining Leibniz's methodological principle, in other words, seems to require a kind of \emph{strengthening} of relativity: that not only space-time but \emph{events themselves} be described \mbox{as relative.}

This paper is organised as follows. In Section~\ref{s2}, we review Bell's two theorems (from 1964 and 1976) and the implicit and explicit assumptions behind them. In Section~\ref{s3}, we break down Bell's 1976 assumption of \LC\ into more primitive concepts to see how this can be given up while still holding to the existence of a \RCA, as Leibniz's methodology requires. In Section~\ref{sec:LF}, we show that this route, which is the one taken by quantum causal models, does not work when it comes to the LF theorem, because \LC\ is not part of that   theorem. In Section~\ref{sec:deeper}, we break the assumptions down to the deepest principles, to display the options allowed by the theorems we consider. Critically, the LF theorem leaves fewer options, and we conclude in Section \ref{s6} by suggesting that cleaving to Leibniz's principle requires rejecting \AOE.

\section{Bell's Theorem(s)}\label{s2}

Bell's theorem, and the LF theorem, are about quantum violations of different conjunctions of metaphysical assumptions through violations of constraints those assumptions imply for empirical correlations between events in certain experimental scenarios. Following \cite{Causarum}, we start by stating two fundamental assumptions  about events and space-time, typically left implicit in discussions of Bell's theorem. \red In \cite{Causarum}, we separated the various assumptions into ``Axioms'', ``Postulates'' and ``Principles''. The ``Axioms'' were so named because they were left as background assumptions in the statement of some or all of the theorems discussed, with their logical implications left implicit. This terminology is accurate for this section, but the LF theorem explicitly uses one of our Axioms in its formulation. \blk Here and in subsequent definitions, whenever a word appears in \textsc{small caps}, it indicates a term whose meaning may be further specified or modified by other assumptions or principles.

\begin{axiom}[\AOE] Every observed \event\ is an absolute single \event, not relative to anything or anyone.
\label{def:AOE}
\end{axiom}

We called this Axiom ``Macroreality'' in \cite{Causarum}, and it plays a crucial role in the LF theorem, as we will see.

\begin{axiom}[\ST]Every \event\ can be located in a background relativistic space-time, where concepts like past and future light-cone, space-like separation, etc., can be made experimentally well-defined.
\label{def:ST}
\end{axiom}

In \cite{Causarum}, we called this assumption ``Minkowski Space-Time'', but a flat space-time is not strictly required,  only a time-orientable pseudo-Riemannian manifold.

In a Bell-type experiment, the observable events under consideration are choices of measurement settings (which we may label $X,Y$ for the case of two distant parties), and their corresponding outcomes (which we may label $A,B$). We will use the same symbol (e.g., $A$) to refer interchangeably to a variable that ranges over possible values of this outcome and to the event corresponding to a particular outcome (say $A=a$) having been observed. We will thus talk about ``correlations between events $A$ and $B$'' as a shorthand for ``correlations between the variables associated with events $A$ and $B$''.

The second of the two assumptions above, \ST, implies that those types of events can be located in space-time to sufficient precision in order to ascertain, e.g., that a pair of events $(X,A)$ is contained in a space-like region separated from a region containing a pair of events $(Y,B)$. The first assumption, \AOE, implies that these variables take well-defined values in every experimental run, and that it is therefore possible to define a conditional probability $p(A,B|X,Y)$ for those events (which in quantum mechanics will be given by the Born rule).

Bell's theorem demonstrates that certain phenomena predicted by quantum mechanics (i.e., the quantum predictions for certain sets of $p(A,B|X,Y)$) cannot be explained by models simultaneously satisfying certain sets of metaphysical assumptions. In our notation, Bell's 1964 theorem can be expressed as: 
\begin{thm}[Bell's 1964 theorem] 
Quantum phenomena violate the conjunction of \NSD, \LOC, and \PRED\ (together with Axioms \ref{def:AOE} and \ref{def:ST}).\label{theorem-Bell64}
\end{thm}

These notions are precisely defined as follows.

\begin{prin}[\NSD]
Any set of \events\ on a space-like hypersurface (SLH) S can be taken to be uncorrelated with any set of interventions subsequent to S.
\end{prin}

This is a rigorous definition for the loose concept sometimes called ``Freedom of Choice'', and what we call ``interventions'' here are usually called ``free choices''. We prefer the term ``intervention'' here as it has a more precise meaning in the literature on causality~\cite{Pearl2000,Hausman1999}. That is, we emphasise that an intervention does not require the free will of a human agent, nor does it need to be itself entirely uncaused. The only important requirement is that an intervention can be chosen via external variables not a priori causally related with any of the other variables relevant to the experiment at hand.
\begin{prin}[\LOC]
The probability of an observable \event\ $e$ is unchanged by conditioning on a space-like-separated intervention $z$, even if it is already conditioned on other \events\ not in the future light-cone of $z$. 
\end{prin}
This is a rigorous version of the concept called ``Parameter Independence'' by Shimony, also discussed (in the same year, and with much the same motivation) by Jarrett~\cite{Jarrett1984}, who called it, as here, ``Locality''. Bell's use of this term in their 1964 theorem~\cite{Bell1964} also accords with this definition~\cite{Wiseman2014}.

\begin{prin}[\PRED]
Any observable \event\ $e$ is determined by a sufficient specification of \events\ on any SLH $S$ prior to $e$, possibly in conjunction with interventions subsequent to S.
\end{prin}
Here we again use the abstract noun employed by Bell in their 1964 paper~\cite{Bell1964}, but note that ``Determinism'' is often also used for the same concept. It is also similar to the concept of ``Outcome Determinism'' used by Spekkens~\cite{Spekkens2005} in relation to contextuality, \red the main difference being that in discussions of contextuality relativistic space-time concepts typically do not play a role.\blk 

\red 
As an aside: here we are using definitions based on those in our earlier work~\cite{Causarum}, with changes in terminology or definitions noted and motivated when they arise, from Section~\ref{s3} onwards. With regard the current section, we have realised that the definitions for \NSD\ and \PRED\ do not match well  with  natural language expectations when applied to models that require a preferred foliation, such as Bohm's~\cite{Bohm1,Bohm2}. We will address that problem in a future publication. However, we note here that: (i) the theorems using these assumptions are still valid; (ii) \PRED\ plays no part in the (more interesting) version of Bell's theorem, from 1976, nor in the LF theorem; (iii) both \NSD\ and \LOC\ are replaceable by the single more natural assumption of \LA, as discussed in Section~\ref{sec:deeper}. \blk 

\begin{figure}
    \centering
    \includegraphics[width=\columnwidth]{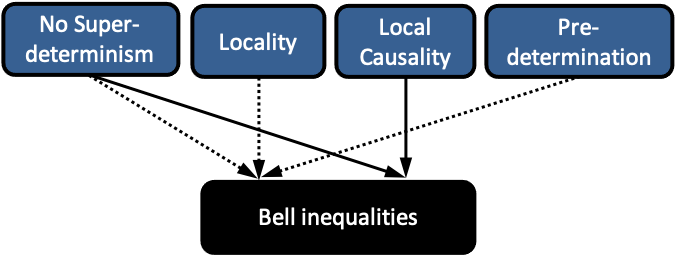}
    \caption{A graphical representation of Bell's 1964 and 1976 theorems. Here and in subsequent figures, a concept is logically implied by the conjunction of its ``parents'' in the graph, i.e., all of those that have arrows pointing to it. A black box represents a concept that is known to be false (i.e., violated in nature). Bell's 1964 theorem is indicated by the dotted arrows and Bell's 1976 theorem by the full arrows.}
    \label{fig:Bells}
\end{figure}
The celebrated Bell experiments of Aspect and {co-workers} 
 \cite{Aspect81,Aspect82} in the early 1980s 
led to almost universal acceptance of the veracity of the phenomena referred to in Bell's theorem \red (theorem~\ref {theorem-Bell64}). \blk 
One of the initial {responses}~\cite{Rohrlich1983} to this was to advocate giving up \PRED\ (or ``hidden variables''~\cite{Rohrlich1983}), by pointing out that ``standard quantum mechanics'' was, after all, an indeterministic theory. Many still hold to this simple argument today. As far as Bell's 1964 theorem alone is concerned, this could allow one to keep \LOC, with the hope of thereby maintaining compatibility with relativity. This was, however, not satisfactory for Bell~\cite{Bell1990}: 

\begin{quote}
``Do we then have to fall back on “no signalling faster than light" as the expression of the fundamental causal structure of contemporary theoretical physics? That is hard for me to accept. For one thing we have lost the idea that correlations can be explained, or at least this idea awaits reformulation''. 
\end{quote} 

As Bell proved in 1976, the Bell inequalities can be derived from a principle specifying (what Bell took to be) a necessary condition for causal explanation in a relativistic space-time, the principle of \LC, which we reformulate as:
\begin{prin}[\LC]
If two space-like separated sets of \events\ {\bf A} and {\bf B} are correlated, then there is a set of \events\ {\bf C} in the intersection of their past light cones such that conditioning on {\bf C} eliminates the correlation.
\end{prin}

With this definition, Bell's 1976 theorem can be formulated as follows.
\begin{thm}[Bell's 1976 theorem]
Quantum phenomena violate the conjunction of \NSD\ and \LC\ (together with Axioms \ref{def:AOE} and \ref{def:ST}).
\end{thm}

The logical implications of Bell's 1964 and 1976 theorems are illustrated in Figure~\ref{fig:Bells}. In light of Bell's 1976 theorem, most physicists conclude (for different reasons, as we will discuss later) that it is Bell's notion of \LC\ that needs to be rejected. \red That said, \blk there are research programs pursuing theories violating \NSD, \red and we note that the assumption of \NSD\ given here can be violated by  retrocausal~\cite{Price2008,Wharton2020}---as well as self-identifiedly  superdeterministic~\cite{Hossenfelder2020,Sen2020}---approaches.\blk

However, if \LC\ is rejected, how can we make sense of causal explanation of correlations in a relativistic space-time? We consider this question in the \mbox{next section}.

\section{Classical and Quantum Causal Explanation}\label{s3}

One of the basic principles of classical causal explanation was proposed by Hans Reichenbach in 1956. We reformulate it below, following~\cite{Cavalcanti2014, Causarum}.

\begin{prin}[\RPCC\ (RPCC)]
If two sets of \events\ {\bf A} and {\bf B} are correlated, and no \event\ in either is a \cause\ of any \event\ in the other, then they have a set of common \causes\ {\bf C}, such that conditioning on {\bf C} eliminates the correlation.
\end{prin}

\begin{figure}
    \centering
    \includegraphics[width=0.75\columnwidth]{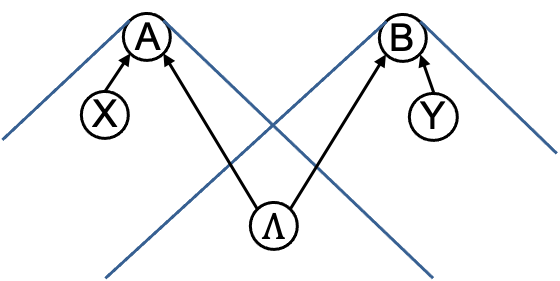}
    \caption{A directed acyclic graph (DAG) representing the causal structure of a Bell scenario involving two parties, Alice and Bob. $X$ and $Y$ represent Alice's and Bob's respective choices of setting, with $A$ and $B$ the corresponding outcomes. $\Lambda$ represents a complete specification of the common causes between the two arms of the experiment. The DAG is motivated by the space-like separation between $(X,A)$ and $(Y,B)$, as indicated by the past light cones of $A$ and $B$.}
    \label{fig:DAG_Bell}
\end{figure}

In more modern terms, Reichenbach's Principle follows from a general framework of (classical) causal models~\cite{Pearl2000} \red (An introduction to the classical causal model framework applied to quantum foundations can be found in \cite{Wood2015}.) \blk In this framework, causal structure is represented by a directed acyclic graph (DAG), with random variables of interest associated with nodes, and arrows between nodes representing an asymmetric cause--effect relationship. The added requirement of acyclicity is intended to exclude causal loops. For example, the causal structure used in the derivation of a Bell inequality in a bipartite scenario has the form shown in Figure~\ref{fig:DAG_Bell}.

In a classical causal model, any probability distribution over the node variables that is compatible with a given graph must satisfy a constraint called the \emph{Causal Markov Condition} (CMC). This can be expressed as the requirement that a variable $X$ is independent of its non-effects $\mathrm{Nd}(X)$ (any nodes that are not among its ``descendants'' in the graph), conditional on its direct causes $\mathrm{Pa}(X)$ (its ``parent'' nodes). That is, 
\begin{equation}
    p(X|\mathrm{Nd}(X),\mathrm{Pa}(X)) = p(X|\mathrm{Pa}(X)) \,.
\end{equation}

The Causal Markov Condition implies \RPCC\ as a special case~\cite{Hitchcock2020}. \red The converse, however, does not hold. To see this, consider a causal graph with three nodes in a chain, $X\rightarrow Y\rightarrow Z$. The CMC implies that the middle node screens off the end nodes, i.e., $p(Z|Y,X)=p(Z|Y)$. RPCC does not imply this condition.\blk 

The classical causal model framework (as the formulation of RPCC above) does not require any a priori assumptions about the causal structure in a Bell scenario. It allows, for example, that the causal structure may include a direct causal connection between a choice of setting (such as $X$ in Figure~\ref{fig:DAG_Bell}) and a space-like separated outcome (such as $B$). To obtain \LC~from this framework, it must be supplemented by some principle relating space-time and causal structure. It is easy to see that the following principle, taken in conjunction with RPCC, is sufficient to imply \LC:
\begin{prin}[\RCA]
Any \causes\ of an \event\ are in its past light-cone.
\end{prin}
\red
We note that this concept was not explicitly defined in ref.~\cite{Causarum} but, as will be discussed in Section~\ref{sec:deeper}, it is a simple consequence of deeper principles introduced there.\blk 

\begin{figure}
    \centering
    \includegraphics[width=\columnwidth]{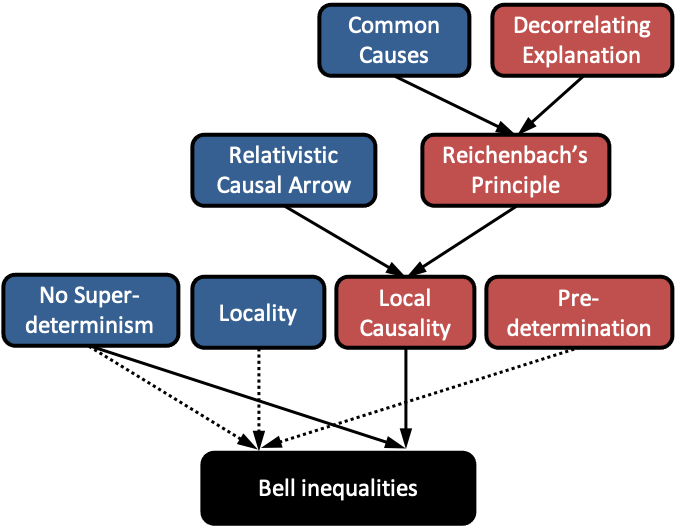}
    \caption{Local Causality is the conjunction of Relativistic Causal Arrow and Reichenbach's principle. As discussed in the text, Leibniz's Principle suggests the rejection Reichenbach's principle, which is implied by the Principle of Common Cause and the Principle of Decorrelating Explanation. Quantum causal models resolve Bell's theorem by rejecting all and only the concepts in red.
    }
    \label{fig:Reichenbach}
\end{figure}

The relationships between the various concepts defined so far are represented in Figure~\ref{fig:Reichenbach}. Thus, if in light of Bell's 1976 theorem, one chooses to reject \LC, one is faced with a dilemma: reject \RCA\ or \RPCC.

\subsection{Leibniz's Principle and Causal Faithfulness}\label{sec:Leibniz}

If one wishes to take the route of rejecting RPCC, the question becomes how to resolve the problem raised by Bell, that ``we have lost the idea that correlations can be explained''. Why not give up \RCA\ instead, as is done, e.g., in Bohmian mechanics~\cite{Bohm1,Bohm2,Maudlin1994}? The reason why theories of this kind remain unattractive to a majority of physicists, we suggest, is that to maintain agreement with observations, they must necessarily violate, at a fundamentally hidden level, some of the operational symmetries we observe in the world. \red (This is not true if one considers a ``non-equilibrium'' version of Bohmian mechanics, which makes predictions different from those of operational quantum theory, as hypothesised by Valentini~\cite{Valentini2010}.)\blk 

More formally, we can say that theories rejecting \RCA\ fall foul of a principle that can be traced back to Leibniz---\emph{Leibniz's Principle of the Identity of Indiscernibles~}\cite{Spekkens2019}---according to which empirically indistinguishable scenarios should be represented by ontologically identical models. This is not to be thought of as a physical principle, but a methodological principle, like Occam's razor, or a form of inference to the best explanation. In choosing between two empirically equivalent theories, we should prefer one that satisfies Leibniz's Principle. 

Spekkens recently argued~\cite{Spekkens2019} that Leibniz's Principle can be identified as the guiding methodological principle underlying Einstein's rationale for both the principle of relativity and the equivalence principle. In other words, to maintain the spirit of the theory of relativity, according to Spekkens, one should look for a theory that satisfies Leibniz's Principle. However, how can it be done, if at all, in light of Bell's theorem? 

Firstly, let us consider: could there be a formulation of quantum theory that satisfies Leibniz's Principle and \RPCC? An obvious candidate would be to reject \NSD\----but can a theory of this type maintain Leibniz's principle?

If the purpose of rejecting \NSD\ is to provide a locally \emph{causal} explanation, then presumably one wants to maintain not only \RPCC, but the entire framework of classical causal models (otherwise the challenge becomes to replace that framework by some alternative). 

Within that framework, one of us has argued~\cite{Cavalcanti2018} that Leibniz's principle implies the principle of \NFT, or {Faithfulness}. This principle requires that every conditional independence between variables (e.g., the no-signalling conditions) must arise as a consequence of the causal graph and not due to special choices of parameters in the model. In particular, therefore, if we cannot operationally signal faster than light, Leibniz's principle suggests that we should prefer a theory that does not postulate faster-than-light-causation.

So could there be some faithful causal structure for quantum correlations, even if perhaps not the one implied by relativity? Unfortunately, as was shown in~\cite{Wood2015}, no classical causal model can explain all instances of bipartite Bell inequality violations while satisfying Faithfulness. This includes even theories violating \NSD, as long as they maintain the DAG structure of classical causal models and the Causal Markov Condition. 

More recently, this result was generalised to arbitrary bipartite~\cite{Cavalcanti2018} and \mbox{multipartite~\cite{Pearl2019}} Bell correlations, demonstrating that the relationship between fine-tuning and Bell nonlocality is generic and not an artefact of the simplest scenarios. \red (It was also shown,  in~\cite{Cavalcanti2018,Pearl2019}, that no classical causal model can explain violations of Kochen-Specker {noncontextuality}~\cite{KS1967} inequalities without fine-tuning. However, as discussed in~\cite{Pearl2019}, this requires a stronger notion of No Fine-Tuning, implicit in \cite{Cavalcanti2018}, but which is also motivated by Leibniz's Principle.)

As argued in Refs.~\cite{Cavalcanti2018,Pearl2019} (and similarly in Ref.~\cite{Spekkens2019}), \blk these results suggest that to maintain Leibniz's Principle, one must reject some of the basic assumptions of the framework of classical causal models, such as \RPCC.

\subsection{``Peaceful Coexistence'' through Quantum Causal Models?}\label{sec:peaceful}

As mentioned in the Introduction, Shimony's proposal for ``peaceful coexistence'' did not satisfy Bell's concerns about causal explanation. Drawing on the causal concepts discussed here so far, the difficulty is that, assuming the standard relativistic causal structure in Figure~\ref{fig:DAG_Bell}, both ``Parameter Independence'' and ``Outcome Independence'' follow from \RPCC.

In 2014, one of us and Lal~\cite{Cavalcanti2014} proposed to break Reichenbach's Principle into two independent principles, which were formalised in Ref.~\cite{Causarum} as 
\begin{prin}[\PCC]
If two sets of \events\ ${\bf A}$ and ${\bf B}$ are correlated, and no \event\ in either is a \cause\ of any \event\ in the other, then they have a set of common \causes\ ${\bf C}$ that {\sc explains} the correlation. 
\end{prin}

\begin{prin}[\DE]
A set of \causes\ ${\bf C}$, common to two sets of events ${\bf A}$ and ${\bf B}$, {\sc explains} a correlation between them only if conditioning on ${\bf C}$ eliminates \mbox{the correlation.}
\end{prin}

Having made this division, we can ask: is it possible to maintain the \PCC\ and replace \DE\ (called ``Factorization of Probabilities'' in~\cite{Cavalcanti2014}) by some other principle of causal explanation in physics? In \cite{Cavalcanti2014}, it was suggested, following the framework of quantum conditional states of Leifer and Spekkens~\cite{Leifer2013}, that a candidate for such principle was the factorization of the Choi-Jamiolkowski operators corresponding to the quantum channels from the common cause $C$ to Alice's and Bob's labs. 

This suggestion was followed, in somewhat different ways, by various proposals for frameworks of quantum causal models~\cite{Pienaar2015,Costa2016,Allen2017,Barrett2019}, generalising the classical causal model formalism~\cite{Pearl2000}. These proposals maintain the ``qualitative'' \red (in the terminology of~\cite{Allen2017}) \blk part of \RPCC\ (the \PCC\ above) while substituting the ``quantitative'' part (\DE) by a suitable quantum generalisation thereof, to arrive at a quantum generalisation of \RPCC. 

This approach to resolve the conflict between relativity and quantum theory is analogous to the route for ``peaceful coexistence'' proposed by Shimony~\cite{Shimony1984}, in that, when applied to a Bell scenario, quantum causal models satisfy Parameter Independence but not Outcome Independence. Nevertheless, unlike Shimony's early proposal, the framework of quantum causal models meets Bell's challenge for providing a (generalised notion of) causal explanation for quantum correlations. It also explains the usefulness of the classical framework, to which it reduces in the appropriate limits~\cite{Barrett2019}. Furthermore, it can provide a faithful causal explanation of Bell correlations~\cite{Costa2016} and allows for causal discovery~\cite{Costa2016}.

This is an interesting program, and it is making steps towards resolving what one of us called the ``easy problem of Bell''~\cite{Cavalcanti2016}, i.e., the problem of giving a causal explanation of Bell correlations. However, as previously argued by one of us in~\cite{Cavalcanti2016}, quantum causal models (as currently formulated) cannot resolve the ``hard problem of Bell'', namely the measurement problem. In the next Section~\ref{sec:LF}, we provide a proof of this assertion, based on the Local Friendliness theorem~\cite{CQD2020}.

\section{The Local Friendliness Theorem}\label{sec:LF}

Wigner's friend paradox is the quintessential intuition pump for the measurement problem. Some recent results have proposed no-go theorems based on extensions of the WFS including multiple observers and entanglement, such as the work by Frauchiger and Renner~\cite{FR2018}. However, unlike Bell's theorem, the Frauchiger-Renner theorem is not theory-independent---that is, it makes assumptions specifically related to quantum theory. \red (As an aside: more recent work generalises that theorem~\cite{Nurgalieva2019}, highlighting the modal-logical nature of some of its assumptions. It could be perhaps said that the Frauchiger--Renner theorem is to the Kochen--Specker {theorem}~\cite{KS1967} as the LF theorem is to Bell's theorem, in the sense that the first two are expressed in terms of assumptions about (modal) logic, whereas the latter two are expressed in terms of metaphysical assumptions.) \blk The Local Friendliness no-go theorem, \red by contrast, {\em} is theory-independent. It \blk is based on the Extended Wigner's Friend scenario introduced by \v{C}aslav \mbox{Brukner \cite{Brukner2017,Brukner2018}.} 

\begin{figure}
    \centering
    \includegraphics[width=0.55\columnwidth]{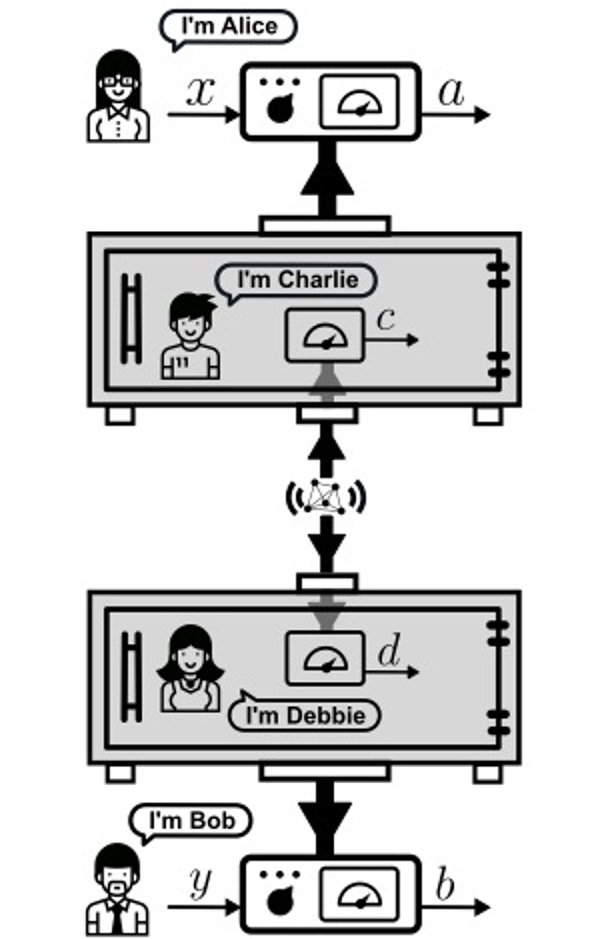}
    \caption{The Extended Wigner's Friend Scenario, reproduced from Ref.~\cite{CQD2020}. See text for details.}
    \label{fig:EWFS}
\end{figure}

In this scenario, depicted in Figure~\ref{fig:EWFS}, there are two labs, controlled by Alice and Bob, each  containing a perfectly isolated vault, with a respective friend inside. The friends share a pair of particles prepared in an entangled state, on which they each perform a measurement on a fixed basis, obtaining outcomes $C$ and $D$, respectively. Alice and Bob have choices of measurements labelled by $X$ and $Y$, with respective outcomes $A$ and $B$. 

\begin{figure}
    \centering
    \includegraphics[width=\columnwidth]{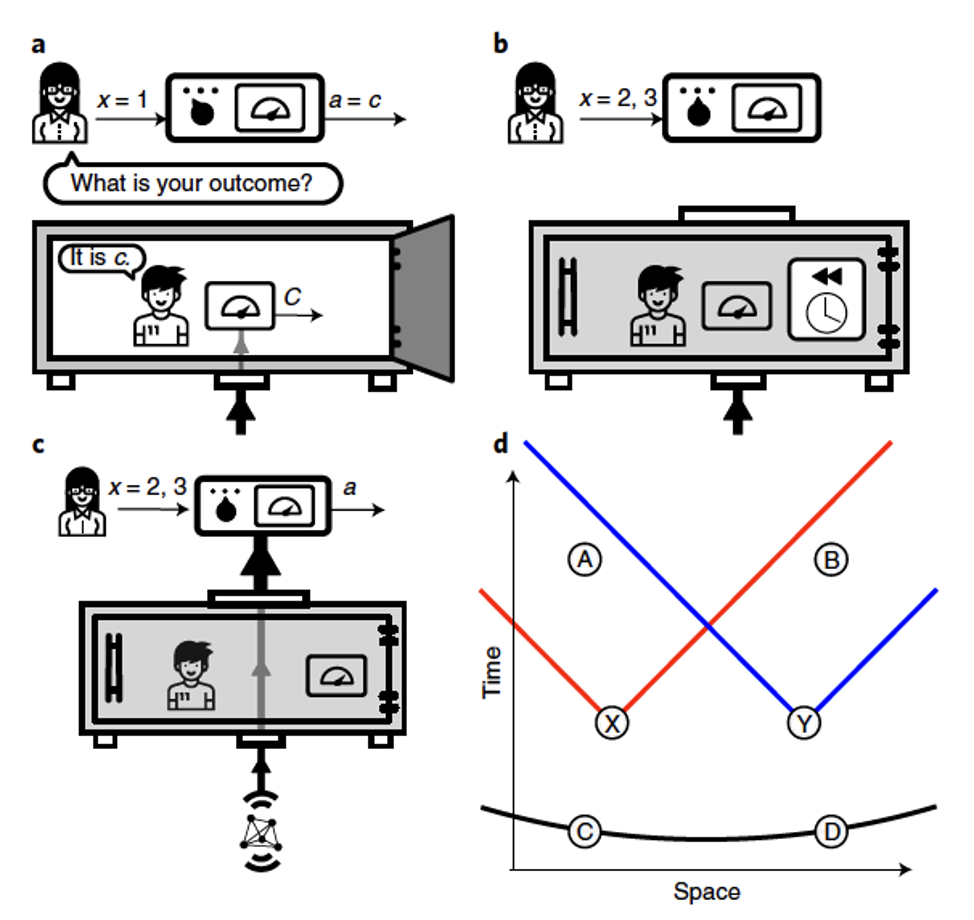}
    \caption{\red 
 {\bf a-c}: The Local Friendliness protocol (see text for details). {\bf d}: A diagram of the space-time locations of the various events involved. (Reproduced from {Ref.}~\cite{CQD2020}).}
    \label{fig:LF_protocol}
\end{figure}

In that protocol, as illustrated in Figure~\ref{fig:LF_protocol}, if Alice chooses $X=1$ \red (Fig.~\ref{fig:LF_protocol}a), \blk she opens the vault and asks Charlie what he observed and sets her own outcome to be equal to that of Charlie, i.e., $A=C$. If $X\neq1$, she performs a measurement on the contents of the vault---including Charlie---in an incompatible basis. This can be done via reversing the unitary evolution that entangled Charlie (and his device, etc.) with his particle \red (Fig.~\ref{fig:LF_protocol}b) \blk and proceeding to perform a measurement on the particle alone, corresponding to a different observable from that observed by Charlie \red (Fig.~\ref{fig:LF_protocol}c). \blk A similar protocol is followed by Bob and Debbie.

For brevity, we will not review all of the details of the LF theorem here but refer the reader to Ref.~\citep{CQD2020}. For the present purposes, the following summary is sufficient. We call \LF\ the conjunction of \LOC, \NSD\ and \AOE. We then show that in an Extended WFS, \red with a space-time arrangement of events as shown in Fig.~\ref{fig:LF_protocol}d, \blk
\LF\ implies constraints on the class of phenomena $p(A,B|X,Y)$ that can be observed by Alice and Bob. \red These constraints can be put \blk in the form of ``LF inequalities''. We then show that LF inequalities can in principle be violated by quantum mechanics, if the requisite quantum operations can in principle be performed on observers. This leads to the following theorem:
\begin{thm}[Local Friendliness no-go theorem]
Quantum phenomena violate the conjunction of \AOE, \LOC\ and \NSD.
\end{thm}

\begin{figure}
    \centering
    \includegraphics[width=0.7\columnwidth]{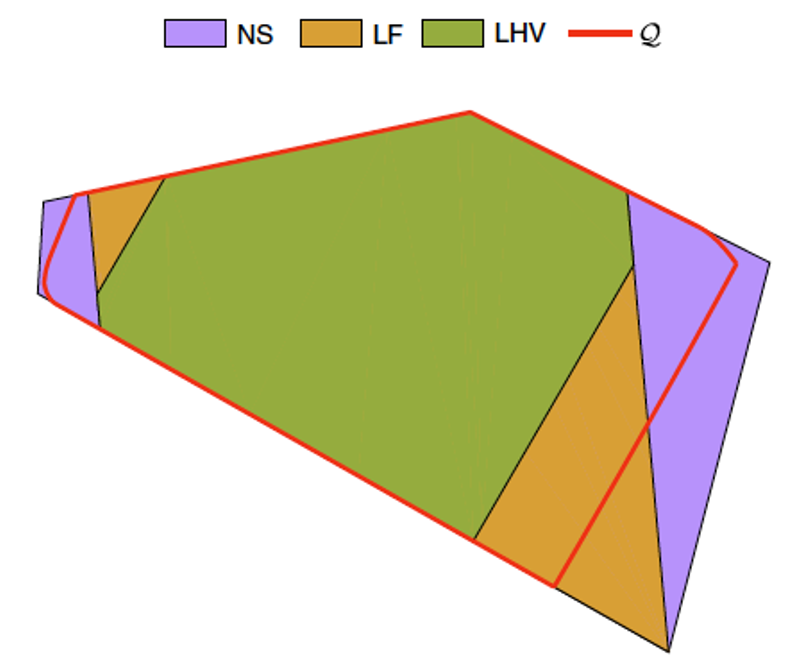}
    \caption{The Local Hidden Variable (LHV)  polytope is the green area, the LF polytope is the orange area, and the purple area is the No-Signalling (NS) polytope. The red line is the boundary of quantum correlations. As the figure shows, the LF inequalities can be violated by quantum mechanics. See~\cite{CQD2020} (from which the figure is reproduced) for further details.}
    \label{fig:sets}
\end{figure}

As depicted in Figure~\ref{fig:sets}, the set of LF correlations (which for a particular scenario has the form of a polytope---the ``LF polytope'') strictly contains the Local Hidden Variable polytope (the set of correlations satisfying the Bell inequalities for a given scenario).  \mbox{In \cite{CQD2020}}, we illustrated this hierarchy in a proof-of-principle experiment involving  polarisation-entangled photons, with the path a photon takes  playing the role of the ``friend'' and a polarising beam splitter the role of the observation.  This hierarchy is a reflection of an important fact: the \LF\ assumptions are strictly weaker than the assumptions needed to derive a Bell inequality. This means that violation of LF inequalities has \emph{strictly stronger implications} than violations of Bell inequalities.

A brief look at Figure~\ref{fig:Reichenbach} shows that it is \emph{not} possible to resolve the LF theorem by dropping \DE\ (and the other assumptions in red), as it was in the ``peaceful coexistence'' resolutions of Bell's theorem---those assumptions are not used in the theorem. As prefigured, this undermines the narrative of the quantum causal models program, and related ideas (such as Shimony's) that quantum mechanics and relativistic causality can be reconciled. 
To better understand what options for reconciliation remain open, we now turn to an  analysis using deeper principles.

\section{Bell and LF Theorems from More Fundamental Causal Principles}
\label{sec:deeper}

In this section, we clarify the implications of the LF theorem for quantum causality, and how it differs from Bell's theorem, by refining many of the concepts introduced so far as consequences of more fundamental principles. Again, we largely follow \cite{Causarum}, with some modifications noted along the way.

We first define the notion of \tsc{causal past} of an \event\ as a set of \events\ containing all of its \causes.
\begin{defin}[\textsc{Causal Past}]
Any \tsc{cause} of an \event\ is in its \tsc{causal past}.
\end{defin}

The next two principles impose spatio-temporal constraints on the \tsc{causal past}.
\begin{prin}[\TCA]
For any \event\ $A$, there is a space-like hypersurface $S$ containing $A$ that separates \events\ in the \tsc{causal past} of $A$ (on the same side of $S$ as $A$'s past light-cone), from \events\ that have $A$ in their \tsc{causal past}.
\end{prin}

This principle is closely related to what we called \textsc{Temporal Order} in \cite{Causarum}, except that here we use \tsc{causal past} instead of \tsc{past}, and further specify that it is in the same temporal direction as the past light-cone, to avoid a potential ambiguity. The following principle is a relaxation of the homonymous principle of Refence~\cite{Causarum}:
\begin{prin}[\RC]
The \tsc{causal past} of an \event\ cannot be outside the light cones of that \event.
\end{prin}

In \cite{Causarum}, the principle of the same name specified that the $\textsc{past}$ is the past light cone, but here the role of picking a direction of time for the $\causes$ is done by \TCA\ instead. \red (The formulation we adopt here seems likely to be useful for disentangling retrocausal and superdeterministic approaches, which we will explore in future work.) \blk In any case, the last two principles imply that the \tsc{causal past} is indeed in the past light cone. Thus, \TCA\ and \RC\ together imply \RCA.

Next we define the principle of \II, a more cautious reformulation of ``Free Choice'' in Ref.~\cite{Causarum} \red (see \cite{Hausman1999} for a detailed discussion of the rationale for this criterion and the role it plays in a manipulability account of causation).\blk 
\begin{prin}[\II]
An intervention has no relevant \causes\; i.e., it can always be chosen via suitable variables that do not have \causes\ among, nor share a common \cause\ with, any of the other experimental variables. 
\end{prin}

Similarly to how \RCA\ arises from the conjunction of two more fundamental principles, 
we also define a principle which arises from the conjunction of \II\ and \PCC. We call this \IC\  (in \citep{Causarum} it was called ``Agent-Causation''): 
\begin{prin}[\IC]
If a set of relevant \events\ $A$ is correlated with an intervention, then that intervention is a \cause\ of at least one \event\ in $A$.
\end{prin}

\begin{figure}
    \centering
    (\textbf{a})\vspace{-2.5mm}
    \includegraphics[width=\columnwidth]{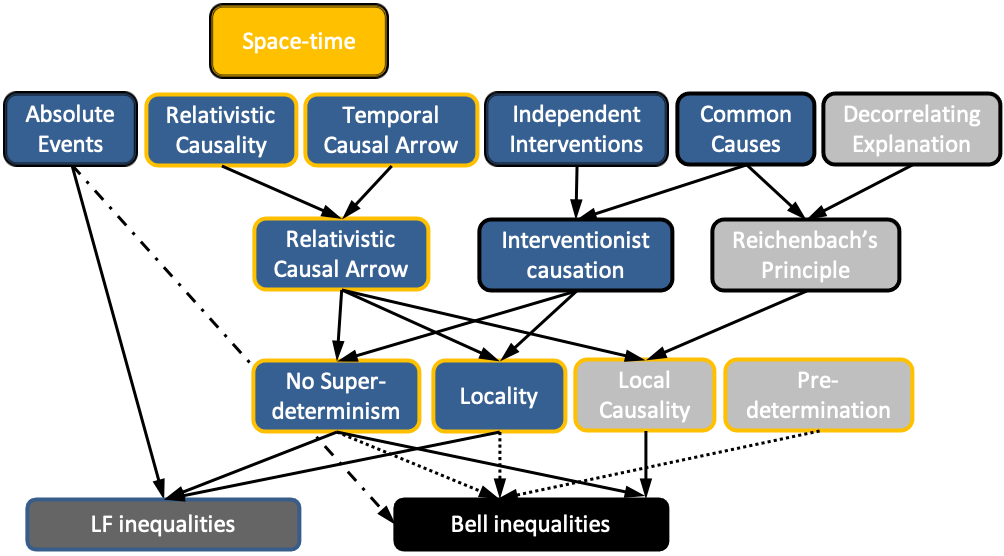}\vspace{7.5mm}
    (\textbf{b})\vspace{-2.5mm}
    \includegraphics[width=\columnwidth]{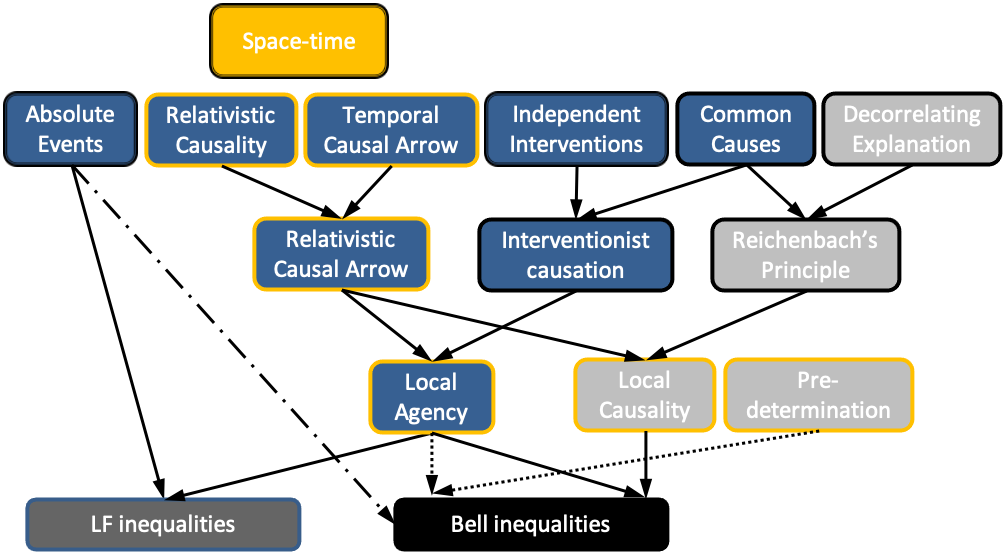}
    \caption{The full set of implications relevant for the Bell and LF theorems. The Bell inequalities can be derived from the conjunction of AOE (the dash-dotted line) together with either the conjunction of the dotted lines (Bell 1964) or the conjunction of the full black lines (Bell 1976). Principles that depend on spatio-temporal concepts assumed in the \ST\ Axiom are represented with a golden outline. In (\textbf{a}), we use the ``traditional'' shallow concepts of \LOC\ and \NSD, whereas in (\textbf{b}), we replace both of these by the deeper concept of \LA. Principles that are rejected within the proposals for ``peaceful coexistence'' between quantum and relativity theories outlined in Section~\ref{sec:peaceful} are represented by light grey boxes. Note that none of those are required to derive LF inequalities.
    }
    \label{fig:Bell-LF}
\end{figure}

We now note that the conjunction of \RCA\ and \IC\ imply both \LOC\ and \NSD, as depicted in Figure~\ref{fig:Bell-LF}a. Indeed, that conjunction also implies the more natural concept {of} \LA~\cite{Causarum}. 

\begin{prin}[\LA]
The only relevant \events\ correlated with an intervention are in its future light cone.
\end{prin}

This can be used in place of \LOC\ and \NSD\, in both Bell's 1964 theorem and the LF theorem, as discussed in Refs.~\cite{Causarum,CQD2020}, respectively. In Bell's 1976 theorem, \LA\ can be used to replace \NSD. All of this is depicted in Fig.~\ref{fig:Bell-LF}b. 

The violation of Bell inequalities thus requires the rejection of at least one of the deeper principles in the top row of Fig.~\ref{fig:Bell-LF}. In the case of the program for ``peaceful coexistence'' discussed in Sec.~\ref{sec:peaceful}, resolution of Bell's theorem is achieved by rejecting \DE, and consequently all of the concepts shown in grey in Fig.~\ref{fig:Bell-LF}, while keeping all the remaining ones. However, the {\bf main point of this paper} is that none of the concepts in grey in Fig.~\ref{fig:Bell-LF} are required for deriving LF inequalities, and thus their rejection is not sufficient to resolve the LF no-go theorem.  Only the blue assumptions go into the LF no-go theorem---in the simplest terms, only \AOE\ and \LA.

\section{Discussion}\label{s6}

Given the conclusion of the preceding section, what is the way forward for the program of quantum causal models?  Firstly, it is important to point out that just as Bell's theorem does not invalidate classical causal models as a useful tool in its regime of applicability, our results do not invalidate quantum causal models as a useful tool in \emph{its} own regime of applicability---namely, in the description of the vast majority of quantum experiments where one can assume, for all practical purposes, a fixed Heisenberg cut---with ``observers'' on one side and ``quantum systems'' on the other. 

One response could thus be to clearly state and accept the limited validity of each of these frameworks, rather than to attempt to resolve the conflict with the LF no-go theorem. This seems somewhat defeatist; a similar response in regards to classical causal models would have precluded the development of quantum causal models. A more interesting response could be to search for a further generalisation of quantum causal models to accommodate Wigner's Friend scenarios.  \blk 

\subsection{ Giving up \LA?}
As we argued earlier in the paper, an underlying motivation for this program is to give a causal explanation for quantum correlations that satisfies Leibniz's principle. Violation of \LA, on the other hand, seems to be in clear violation of this principle, insofar as we are not able to send signals outside of the future light cones of an intervention. However, let us consider whether this may be a too hasty conclusion.

Firstly, the program of quantum causal models has a clear aim of maintaining the \PCC\ as a basic requirement for causal explanation. It also relies on an interventionist notion of causation for effective causal discovery and thus on the principle of \II. It would also seem contrary to the aims of that program to reject \RC\, providing a causal explanation compatible with relativity being one of its main aims. 

\TCA\ is required for obtaining a causal graph that has the form of a DAG, as is assumed in the initial formalisms for quantum causal models. On the other hand, the two most developed of these are based on the process matrix formalism, and as is well known, this formalism allows in general for causally non-separable processes, which are interpreted as representing situations with indefinite causal structure~\cite{Oreshkov2012}. Indeed, recent work~\cite{Barrett2020} considers an extension of the framework of quantum causal models of~\cite{Allen2017,Barrett2019} to allow for cyclic causal graphs and show that this allows for a representation of certain classes of causally nonseparable processes, including some processes that violate causal inequalities as cyclic quantum causal models. However, there is no reason to think that postulating a causally nonseparable process would be sufficient to resolve the LF theorem, as we now explain. 

The process matrix formalism~\cite{Oreshkov2012,Costa2016,Barrett2019} is limited to describing an experimental situation in terms of several labs receiving a quantum system as an input, upon which an instrument (represented as a set of CP maps that sum to a CPTP map) can be performed, the outcome of which (one of these CP maps) is denoted an ``event''. After this CP map is applied to the system, it is sent through an output in the lab into the rest of the world, represented by the ``process'' to be potentially routed to the other labs, possibly in a causally nonseparable way. However, the agents, as well as the devices they use to perform their required instruments and obtain their corresponding outcomes, are left outside of the quantum description given by the process, just as in textbook operational quantum mechanics. In other words, the only thing that leaves the labs in that formalism is the quantum systems being measured by the various agents. In a Wigner's Friend Scenario (WFS), however, \emph{the entire contents of the friend's lab} can be part of the input quantum system for a superobserver. This scenario simply cannot be described in current versions of the process matrix formalism.

Furthermore, one must recall that we require not only the violation of \TCA, but that it is violated in such a way as to allow for the violation \LA, if this is to resolve the LF no-go theorem.

\subsection{ Giving up \AOE?}
The remaining alternative is to give up \AOE. How does that fare as a path forward for the program of quantum causal models? We note from the discussion above that in quantum causal models, an ``event'' is the outcome of an instrument, associated with a CP map, and that the instrument is usually described as a classical variable, with a fixed ``Heisenberg cut'' applied in each lab. In a WFS, a fully quantum description of the ``friends'' and their labs is required instead. This suggests a direction to search for a generalisation of the process matrix formalism where the events observed in each lab can be described as \emph{relative events}, where the outcomes of each instrument are encoded in relational variables associated with each lab, but which may not necessarily take well-defined values from a global perspective encompassing all of the labs. In such a formalism, there would not necessarily be a joint probability distribution over events observed in all labs, as usually assumed in the standard formalism.

In other words, if something like this suggestion is possible, the resolution of the conflict between quantum mechanics and relativity requires a \emph{strengthening} of relativity: that not only space-time but \emph{events themselves} be regarded as relative.

\AOE\ is rejected in some interpretations of quantum mechanics such as {Everett}~\cite{Everett1957}, {Relational QM}~\cite{Rovelli1996} and {QBism}~\cite{Fuchs2012,Fuchs2014a}.

{Everett}~\cite{Everett1957} says (emphasis in the original), {``One can arbitrarily choose a state for one subsystem, and be led to the relative state for the remainder. Thus, we are faced with a fundamental \emph{relativity of states}, which is implied by the formalism of composite systems. It is meaningless to ask the absolute state of a subsystem—one can only ask the state relative to a given state of the remainder of the subsystem''}, and later adds that {``with each succeeding observation (or interaction), the observer state `branches' into a number of different states. Each branch represents a different outcome of the measurement and the corresponding eigenstate for the object-system state.''} 

Relational QM is rooted in the work of Everett, but he does not subscribe to realism about the universal wave function. Instead, according to Laudisa and {Rovelli}~\cite{Laudisa2021}: {``The world is therefore described by RQM as an evolving network of sparse relative \emph{events}, described by punctual relative values of physical variables''}.

QBism takes a more radical position, where a measurement outcome is a personal experience of an observer: {In}~\cite{Fuchs2012}, Fuchs says: {``What we learn from Wigner and their friend is that we all have truly private worlds in addition to our public worlds''}. Furthermore, {in}~\cite{Fuchs2014a}, Mermin and Schack say: {``What is real for an agent rests entirely on what that agent experiences, and different agents have different experiences''}. A detailed discussion of the implications of the Local Friendliness no-go theorem for QBism is given in~\cite{Cavalcanti2021}.

However, these accounts \blk do not give a complete response to the challenge of providing a causal explanation that extends the classical framework. Whether this direction will work remains to be seen, but it certainly opens several questions. For example, if events are not absolute, then what is the meaning of the axiom of Space-Time? Would it need to also be generalised, perhaps by relaxing the assumption that there exists a single background space-time? Furthermore, if the motivation for rejecting \AOE\ is to keep \LA, what is the meaning of the notion of \events\ required in that principle? We conjecture that \LA\ can be maintained at least in a suitably relaxed form, from the perspective of each agent.  However, these are very challenging problems that are far beyond the scope of this paper. We \blk also suggest that a fully satisfactory resolution of the measurement problem, underlying the LF no-go theorem, would require that these concepts either be ultimately explicable without direct reference to agents, or that they can be understood as describing an emergent level of description where these agent-centric concepts are applicable.

Whatever the solution, the implication is that something much more radical than we have been able to conceive so far is required for quantum causality to resolve the measurement problem in the form of the LF no-go theorem. We still need radical \mbox{conceptual renewal.}

\vspace{6pt}


\textit{Acknowledgements.---} 
Both authors contributed to all aspects of the work. The initial draft was based on talks presented by EGC at the "Causality in a quantum world" workshop at Anacapri in Sep. 2019, and at the Harvard Foundations of Physics mini-workshop on Causation in Oct 2020 (online). Figures 1-3 and 7 were made by EGC. This work was supported by the Australian Research Council (ARC) Future Fellowship FT180100317, and grants ``Events, agents, and causation in ontological models of quantum theory'' (FQXi-RFP-1807) and ``Quantum and consciousness: paths to experiment, and implications for interpretations'' (FQXi-RFP-CPW-2019) from the Foundational Questions Institute and Fetzer Franklin Fund, a donor advised fund of Silicon Valley Community Foundation. We acknowledge the traditional owners of the land at at Griffith University on which this work was undertaken, the Yuggera and Yugambeh peoples.


\begin{thebibliography}{999}

\bibitem[Bell(1987)]{Bell1984}
Bell, J.S.
\newblock Speakable and Unspeakable in Quantum Mechanics. {{Introduction}}
  Remarks at {{Naples}}-{{Amalfi}} Meeting, {{May}} 7, 1984. In {\em Speakable
  and Unspeakable in Quantum Mechanics}; Bell, J.S., Ed.; {Cambridge University
  Press}: {Cambridge},  1987; p. 172.

\bibitem[Bong \em{et~al.}(2020)Bong, {Utreras-Alarc{\'o}n}, Ghafari, Liang,
  Tischler, Cavalcanti, Pryde, and Wiseman]{CQD2020}
Bong, K.W.; {Utreras-Alarc{\'o}n}, A.; Ghafari, F.; Liang, Y.C.; Tischler, N.;
  Cavalcanti, E.G.; Pryde, G.J.; Wiseman, H.M.
\newblock A Strong No-Go Theorem on the {{Wigner}}'s Friend Paradox.
\newblock {\em Nature Physics} {\bf 2020}, {\em 16},~1199--1205, \href{https://arxiv.org/abs/1907.05607}{{\normalfont [1907.05607]}}.

\bibitem[Shimony(1984)]{Shimony1984}
Shimony, A.
\newblock Controllable and Uncontrollable Non-Locality. In {\em Foundations of
  {{Quantum Mechanics}} in {{Light}} of {{New Technology}}}; and~others
  Kamefuchi, S., Ed.; {The Physical Society of Japan}: {Tokyo},  1984; pp.
  225--230.

\bibitem[Myrvold(2002)]{Myrvold2002}
Myrvold, W.C.
\newblock On Peaceful Coexistence: Is the Collapse Postulate Incompatible with
  Relativity?
\newblock {\em Studies in History and Philosophy of Science Part B: Studies in
  History and Philosophy of Modern Physics} {\bf 2002}, {\em 33},~435--466.

\bibitem[Berkovitz(2007)]{Berkovitz2007}
Berkovitz, J.
\newblock Action at a {{Distance}} in {{Quantum Mechanics}}.
\newblock {\em The Stanford Encyclopedia of Philosophy} {\bf 2007}.

\bibitem[Bell(1990)]{Bell1990}
Bell, J.S.
\newblock La {{Nouvelle Cuisine}}. In {\em Between {{Science}} and
  {{Technology}}}; {Elsevier B.V.},  1990.

\bibitem[Leifer and Spekkens(2013)]{Leifer2013}
Leifer, M.S.; Spekkens, R.W.
\newblock Towards a Formulation of Quantum Theory as a Causally Neutral Theory
  of {{Bayesian}} Inference.
\newblock {\em Physical Review A} {\bf 2013}, {\em 88},~052130,
  \href{http://www.arxiv.org/abs/1107.5849}{{\normalfont [1107.5849]}}.

\bibitem[Cavalcanti and Lal(2014)]{Cavalcanti2014}
Cavalcanti, E.G.; Lal, R.
\newblock On Modifications of {{Reichenbach's}} Principle of Common Cause in
  Light of {{Bell's}} Theorem.
\newblock {\em Journal of Physics A: Mathematical and Theoretical} {\bf 2014},
  {\em 47},~424018,  \href{http://arxiv.org/abs/1311.6852}{{\normalfont
  [1311.6852]}}.

\bibitem[Pienaar and Brukner(2015)]{Pienaar2015}
Pienaar, J.; Brukner, C.
\newblock A Graph-Separation Theorem for Quantum Causal Models.
\newblock {\em New Journal of Physics} {\bf 2015}, {\em 17},~073020,
  \href{http://arxiv.org/abs/1406.0430}{{\normalfont [1406.0430]}}.

\bibitem[Costa and Shrapnel(2016)]{Costa2016}
Costa, F.; Shrapnel, S.
\newblock Quantum Causal Modelling.
\newblock {\em New Journal of Physics} {\bf 2016}, {\em 18},~063032,
  \href{http://arxiv.org/abs/1512.07106}{{\normalfont [1512.07106]}}.

\bibitem[Allen \em{et~al.}(2017)Allen, Barrett, Horsman, Lee, and
  Spekkens]{Allen2017}
Allen, J.M.A.; Barrett, J.; Horsman, D.C.; Lee, C.M.; Spekkens, R.W.
\newblock Quantum {{Common Causes}} and {{Quantum Causal Models}}.
\newblock {\em Physical Review X} {\bf 2017}, {\em 7},~031021,
  \href{http://arxiv.org/abs/1609.09487}{{\normalfont [1609.09487]}}.

\bibitem[Barrett \em{et~al.}(2019)Barrett, Lorenz, and Oreshkov]{Barrett2019}
Barrett, J.; Lorenz, R.; Oreshkov, O.
\newblock Quantum {{Causal Models}} {\bf 2019}.
\newblock  \href{http://arxiv.org/abs/1906.10726}{{\normalfont
  [1906.10726]}}.

\bibitem[Pearl(2000)]{Pearl2000}
Pearl, J.
\newblock {\em Causality: {{Models}}, {{Reasoning}} and {{Inference}}};
  {Cambridge University Press},  2000.

\bibitem[Wiseman and Cavalcanti(2017)]{Causarum}
Wiseman, H.M.; Cavalcanti, E.G.
\newblock Causarum {{Investigatio}} and the {{Two Bell}}'s {{Theorems}} of
  {{John Bell}}. In {\em Quantum [{{Un}}]{{Speakables II}} - {{Half}} a
  {{Century}} of {{Bell}}'s {{Theorem}}}; Bertlmann, R.; Zeilinger, A., Eds.;
  {Springer International Publishing},  2017; pp. 119--142,
  \href{http://arxiv.org/abs/1503.06413}{{\normalfont [1503.06413]}}.

\bibitem[Bell(1964)]{Bell1964}
Bell, J.S.
\newblock On the {{Einstein}}-{{Podolsky}}-{{Rosen Paradox}}.
\newblock {\em Physics} {\bf 1964}, {\em 1},~195.

\bibitem[Bell(1976)]{Bell1976}
Bell, J.S.
\newblock The Theory of Local Beables.
\newblock {\em Epistemological Letters} {\bf 1976}, {\em 9},~11--24.

\bibitem[Wiseman(2014)]{Wiseman2014}
Wiseman, H.M.
\newblock The Two {{Bell}}'s Theorems of {{John Bell}}.
\newblock {\em J. Phys. A: Math. Theor.} {\bf 2014}, {\em 47},~424001.

\bibitem[Wiseman \em{et~al.}(2021)Wiseman, Cavalcanti, and Rieffel]{LF_AI}
Wiseman, H.M.; Cavalcanti, E.G.; Rieffel, E.G.
\newblock A thoughtful ``Local Friendliness'' no-go theorem.
\newblock In preparation.

\bibitem[Proietti \em{et~al.}(2019)Proietti, Pickston, Graffitti, Barrow,
  Kundys, Branciard, Ringbauer, and Fedrizzi]{Proietti2019}
Proietti, M.; Pickston, A.; Graffitti, F.; Barrow, P.; Kundys, D.; Branciard,
  C.; Ringbauer, M.; Fedrizzi, A.
\newblock Experimental Test of Local Observer Independence.
\newblock {\em Sci Adv} {\bf 2019}, {\em 5},~eaaw9832.

\bibitem[Spekkens(2019)]{Spekkens2019}
Spekkens, R.W.
\newblock The Ontological Identity of Empirical Indiscernibles: {{Leibniz}}'s
  Methodological Principle and Its Significance in the Work of {{Einstein}}
  {\bf 2019}.
\newblock  \href{http://arxiv.org/abs/1909.04628}{{\normalfont
  [1909.04628]}}.

\bibitem[Bohm(1952{\natexlab{a}})]{Bohm1}
Bohm, D.
\newblock A {{Suggested Interpretation}} of the {{Quantum Theory}} in {{Terms}}
  of "{{Hidden}}" {{Variables}}. {{I}}.
\newblock {\em Physical Review} {\bf 1952}, {\em 85},~166--179.

\bibitem[Bohm(1952{\natexlab{b}})]{Bohm2}
Bohm, D.
\newblock A {{Suggested Interpretation}} of the {{Quantum Theory}} in {{Terms}}
  of "{{Hidden}}" {{Variables}}. {{II}}.
\newblock {\em Physical Review} {\bf 1952}, {\em 85},~180--193.

\bibitem[Hausman and Woodward(1999)]{Hausman1999}
Hausman, D.M.; Woodward, J.
\newblock Independence, Invariance and the Causal {{Markov}} Condition.
\newblock {\em Br J Philos Sci} {\bf 1999}, {\em 50},~521--583.

\bibitem[Jarrett(1984)]{Jarrett1984}
Jarrett, J.P.
\newblock On the Physical Significance of the Locality Conditions in the
  {{Bell}}-Arguments.
\newblock {\em Nous} {\bf 1984}, {\em 18},~569--589.

\bibitem[Spekkens(2005)]{Spekkens2005}
Spekkens, R.W.
\newblock Contextuality for Preparations, Transformations, and Unsharp
  Measurements.
\newblock {\em Physical Review A} {\bf 2005}, {\em 71},~052108,
  \href{http://arxiv.org/abs/quant-ph/0406166}{{\normalfont
  [quant-ph/0406166]}}.

\bibitem[Aspect \em{et~al.}(1981)Aspect, Grangier, and Roger]{Aspect81}
Aspect, A.; Grangier, P.; Roger, G.
\newblock Experimental Tests of Realistic Local Theories via {Bell}'s Theorem.
\newblock {\em Phys. Rev. Lett.} {\bf 1981}, {\em 47},~460--463.
\newblock
  doi:\href{https://doi.org/10.1103/PhysRevLett.47.460}{10.1103/PhysRevLett.47.460}.
  
\bibitem[Aspect \em{et~al.}(1982)Aspect, Dalibard, and Roger]{Aspect82}
Aspect, A.; Dalibard, J.; Roger, G.
\newblock Experimental Test of {Bell}'s Inequalities Using Time- Varying
  Analyzers.
\newblock {\em Phys. Rev. Lett.} {\bf 1982}, {\em 49},~1804--1807.
\newblock
  doi:\href{https://doi.org/10.1103/PhysRevLett.49.1804}{10.1103/PhysRevLett.49.1804}.

\bibitem[Rohrlich(1983)]{Rohrlich1983}
Rohrlich, F.
\newblock Facing Quantum Mechanical Reality.
\newblock {\em Science} {\bf 1983}, {\em 221},~1251--1255.
\newblock
  doi:\href{https://doi.org/10.1126/science.221.4617.1251}{\detokenize{10.1126/science.221.4617.1251}}.

\bibitem[Hossenfelder and Palmer(2020)]{Hossenfelder2020}
Hossenfelder, S.; Palmer, T.
\newblock Rethinking {{Superdeterminism}}.
\newblock {\em Front. Phys.} {\bf 2020}, {\em 8}.

\bibitem[Sen and Valentini(2020)]{Sen2020}
Sen, I.; Valentini, A.
\newblock Superdeterministic Hidden-Variables Models {{I}}: Nonequilibrium and
  Signalling.
\newblock Technical report,  2020,
  \href{http://arxiv.org/abs/2003.11989}{{\normalfont [2003.11989]}}.

\bibitem[Price(2008)]{Price2008}
Price, H.
\newblock Toy Models for Retrocausality.
\newblock {\em Studies in History and Philosophy of Science Part B: Studies in
  History and Philosophy of Modern Physics} {\bf 2008}, {\em 39},~752--761,
  \href{http://arxiv.org/abs/0802.3230}{{\normalfont [0802.3230]}}.

\bibitem[Wharton and Argaman(2020)]{Wharton2020}
Wharton, K.B.; Argaman, N.
\newblock Colloquium: {{Bell}}'s Theorem and Locally Mediated Reformulations of
  Quantum Mechanics.
\newblock {\em Rev. Mod. Phys.} {\bf 2020}, {\em 92},~021002.

\bibitem[Wood and Spekkens(2015)]{Wood2015}
Wood, C.J.; Spekkens, R.W.
\newblock The Lesson of Causal Discovery Algorithms for Quantum Correlations:
  Causal Explanations of {{Bell}}-Inequality Violations Require Fine-Tuning.
\newblock {\em New Journal of Physics} {\bf 2015}, {\em 17},~033002,
  \href{http://arxiv.org/abs/1208.4119}{{\normalfont [1208.4119]}}.

\bibitem[Hitchcock and R{\'e}dei(2020)]{Hitchcock2020}
Hitchcock, C.; R{\'e}dei, M.
\newblock Reichenbach's {{Common Cause Principle}}.
\newblock {\em The Stanford Encyclopedia of Philosophy} {\bf 2020}.

\bibitem[Maudlin(1994)]{Maudlin1994}
Maudlin, T.
\newblock {\em Quantum {{Non}}-Locality and {{Relativity}}}; {Blackwell},
  1994.

\bibitem[Valentini(2010)]{Valentini2010}
Valentini, A.
\newblock Beyond the {{Quantum}}.
\newblock {\em Physics World} {\bf 2010}, {\em 22},~32--37,
  \href{http://arxiv.org/abs/1001.2758}{{\normalfont [1001.2758]}}.

\bibitem[Cavalcanti(2018)]{Cavalcanti2018}
Cavalcanti, E.G.
\newblock Classical {{Causal Models}} for {{Bell}} and {{Kochen}}-{{Specker
  Inequality Violations Require Fine}}-{{Tuning}}.
\newblock {\em Physical Review X} {\bf 2018}, {\em 8},~021018.
\newblock  \href{https://arxiv.org/abs/1705.05961}{{\normalfont
  [1705.05961]}}.


\bibitem[Pearl and Cavalcanti(2019)]{Pearl2019}
Pearl, J.C.; Cavalcanti, E.G.
\newblock Classical Causal Models Cannot Faithfully Explain {{Bell}}
  Nonlocality or {{Kochen}}-{{Specker}} Contextuality in Arbitrary Scenarios
  {\bf 2019}.
\newblock  {\em Quantum} {\bf 2021}, {\em 5}, 518, \href{http://arxiv.org/abs/1909.05434}{{\normalfont
  [1909.05434]}}.

\bibitem[Kochen and Specker(1967)]{KS1967}
Kochen, S.; Specker, E.P.
\newblock The Problem of Hidden Variables in Quantum Mechanics.
\newblock {\em Journal of Mathematics and Mechanics} {\bf 1967}, {\em
  17},~59--87.

\bibitem[Cavalcanti(2016)]{Cavalcanti2016}
Cavalcanti, E.G.
\newblock Bell's Theorem and the Measurement Problem: Reducing Two Mysteries to
  One?
\newblock {\em Journal of Physics: Conference Series} {\bf 2016}, {\em
  701},~012002,  \href{http://arxiv.org/abs/1602.07404}{{\normalfont
  [1602.07404]}}.

\bibitem[Frauchiger and Renner(2018)]{FR2018}
Frauchiger, D.; Renner, R.
\newblock Quantum Theory Cannot Consistently Describe the Use of Itself.
\newblock {\em Nature Communications} {\bf 2018}, {\em 9},~3711.

\bibitem[Nurgalieva and {del Rio}(2019)]{Nurgalieva2019}
Nurgalieva, N.; {del Rio}, L.
\newblock Inadequacy of {{Modal Logic}} in {{Quantum Settings}}.
\newblock {\em Electron. Proc. Theor. Comput. Sci.} {\bf 2019}, {\em
  287},~267--297.

\bibitem[Brukner(2017)]{Brukner2017}
Brukner, C.
\newblock On the Quantum Measurement Problem. In {\em Quantum
  [{{Un}}]{{Speakables II}} - {{Half}} a {{Century}} of {{Bell}}'s
  {{Theorem}}}; Bertlmann, R.; Zeilinger, A., Eds.; {Springer International
  Publishing},  2017; pp. 95--117,
  \href{http://arxiv.org/abs/1507.05255}{{\normalfont [1507.05255]}}.

\bibitem[Brukner(2018)]{Brukner2018}
Brukner, C.
\newblock A {{No}}-{{Go Theorem}} for {{Observer}}-{{Independent Facts}}.
\newblock {\em Entropy} {\bf 2018}, {\em 20},~350,
  \href{http://arxiv.org/abs/1804.00749}{{\normalfont [1804.00749]}}.

\bibitem[Oreshkov \em{et~al.}(2012)Oreshkov, Costa, and Brukner]{Oreshkov2012}
Oreshkov, O.; Costa, F.; Brukner, C.
\newblock Quantum Correlations with No Causal Order.
\newblock {\em Nature Communications} {\bf 2012}, {\em 3},~1092,
  \href{http://arxiv.org/abs/1105.4464}{{\normalfont [1105.4464]}}.

\bibitem[Barrett \em{et~al.}(2020)Barrett, Lorenz, and Oreshkov]{Barrett2020}
Barrett, J.; Lorenz, R.; Oreshkov, O.
\newblock Cyclic {{Quantum Causal Models}} {\bf 2020}.
\newblock  \href{http://arxiv.org/abs/2002.12157}{{\normalfont
  [2002.12157]}}.

\bibitem[Everett(1957)]{Everett1957}
Everett, H.
\newblock {``Relative state'' formulation of quantum mechanics}.
\newblock {\em Rev. Mod. Phys.} {\bf 1957}, {\em 29},~454--462.

\bibitem[Rovelli(1996)]{Rovelli1996}
Rovelli, C.
\newblock {Relational quantum mechanics}.
\newblock {\em International Journal of Theoretical Physics} {\bf 1996}, {\em
  35},~1637--1678.
  
\bibitem[Fuchs(2012)]{Fuchs2012}
Fuchs, C.A.
\newblock Interview with a Quantum Bayesian {\bf 2012}.
\newblock  \href{https://arxiv.org/abs/1207.2141}{{\normalfont
  [1207.2141]}}.

\bibitem[Fuchs(2014a)]{Fuchs2014a}
Fuchs, C.A.; Mermin, N.D.; Schack, R.
\newblock An Introduction to QBism with an Application to the Locality of Quantum Mechanics.
\newblock {\em American Journal of Physics} {\bf 2014}, {\em
  82},~749–754.
  
\bibitem[Laudisa(2021)]{Laudisa2021}
{Laudisa, F.; Rovelli, C.}
\newblock \href{https://plato.stanford.edu/archives/spr2021/entries/qm-relational/}{Relational Quantum Mechanics.}
\newblock {{\em The Stanford Encyclopedia of Philosophy (Spring 2021 Edition)}; Edward N. Zalta, Ed.}.

\bibitem[Cavalcanti(2021)]{Cavalcanti2021}
Cavalcanti, E.G.
\newblock The {{View}} from a {{Wigner Bubble}}.
\newblock {\em Found Phys} {\bf 2021}, {\em 51},~39.
\newblock  \href{https://arxiv.org/abs/2008.05100}{{\normalfont [2008.05100]}}
  
  
\end{thebibliography}
\end{document}